\documentclass[aps,preprint,preprintnumbers,amsmath,amssymb,nofootinbib]{revtex4}
\usepackage{graphicx}
\usepackage{epsfig}
\begin{document}
\title{Particles and gravitons creation after inflation from a 5D vacuum}
\author{$^{1,2}$ Mariano Anabitarte\footnote{E-mail address: anabitar@mdp.edu.ar} and $^{1,2}$ Mauricio Bellini \footnote{ E-mail address:
mbellini@mdp.edu.ar, mbellini@conicet.gov.ar}}

\address{$^1$ Departamento de F\'isica, Facultad de Ciencias Exactas y
Naturales, Universidad Nacional de Mar del Plata, Funes 3350,
C.P. 7600, Mar del Plata, Argentina.\\  \\
$^2$ Instituto de Investigaciones en F\'{\i}sicas de Mar del Plata (IFIMAR), \\
Consejo acional de Investigaciones Cient\'ificas y T\'ecnicas
(CONICET), Argentina.}

\begin{abstract}
We use the Bogoliubov formalism to study both, particles and
gravitons creation at the reheating epoch, after a phase
transition from inflation to a radiation dominated universe. The
modes of the inflaton field fluctuations and the scalar
fluctuations of the metric at the end of inflation are obtained by
using a recently introduced formalism related to the Induced
Matter theory of gravity. The interesting result is that the
number of created particles is bigger than $10^{90}$ on
cosmological scales. Furthermore, the number of gravitons are
nearly $10^{-17}$ times smaller than the number of created
particles. In both cases, these numbers rapidly increase on
cosmological scales.
\end{abstract}
\maketitle

\section{Introduction}

Inflation has become the standard paradigm for explaining the
homogeneity and the isotropy of our observed Universe \cite{1,2}.
During this epoch the energy density of the Universe was dominated
by some scalar field (the inflaton), with negligible kinetic
energy density, such that the corresponding vacuum energy density
was the responsible for the exponential growth of the scale factor
of the universe. During this phase a small and smooth region of
the order of size of the Hubble radius grew so large that it
easily encompassed the comoving volume of the entire presently
observed Universe. This is the reason for which the observed
Universe is so homogeneous and isotropic. Furthermore, it is now
clear that structure in the Universe comes primarily from an
almost scale-invariant super-horizon curvature perturbation. This
perturbation originates presumably from the vacuum fluctuation,
during the almost-exponential inflation, of some field with mass
much less than the Hubble parameter $H_0$. Indeed, any scalar
field whose mass is lighter than $H_0$ suffered, in a (quasi) de
Sitter epoch, with a scale independent quantum fluctuations
spectrum\cite{3,4,5,5b}.

During inflation, particle production can only occur for particles
that are light compared to the Hubble scale without classical
conformal invariance; gravitons and massless minimally coupled
scalars and light fermions are unique in that respect. Particle
creation during inflation was studied many years ago in the
framework of standard inflation\cite{cal,calb}. The process
appeared to be straightforward: in models like new inflation and
chaotic inflation\cite{chao}. These models incorporate a second
order phase transition to end inflation, the inflaton field would
wind up oscillating around the minimum of its potential near the
end of inflation. These oscillations would produce a sea of
relativistic particles, if one added (by hand) interaction terms
between the inflaton and these lighter species. The direct
production from vacuum fluctuations during inflation of X-particle
was considered in\cite{ci}. Particle creation has been also
considered in warm inflationary\cite{wi} and fresh
inflationary\cite{fi} scenarios. More recently, particle creation
during inflation was considered in a model of brane inflation
where two stacks of mobile branes are moving ultra
relativistically in a warped throat\cite{bi}.

After inflation, the universe could have suffered a phase
transition from a de Sitter vacuum dominated state to a
decelerated radiation dominated stage. At this moment, a great
amount of the potential density energy is transferred to radiation
energy density due to the decay of the inflaton field produced by
its interaction with other boson
fields\cite{rad,radb,radc,radd,rade}. Some years ago
Kaiser\cite{kaiser} demonstrated that particles produced from the
parametric resonance effect when $H \neq 0$ is fewer than in the
$H\simeq 0$ case, but can still be exponentially greater than when
the resonance is neglected altogether.

On the other hand, in a previous work\cite{abm} we studied the
scalar metric fluctuations of a 5D spacetime background metric
$({\cal M}, g)$, which is Riemann-flat and hence Ricci-flat:
$\hat{\cal R}_{AB}=0$. From the mathematical point of view, the
Campbell-Magaard
theorem\cite{campbell,campbellb,campbellc,campbelld} serves as a
ladder to go between manifolds whose dimensionality differs by
one. This theorem, which is valid in any number of dimensions,
implies that every solution of the 4D Einstein equations with
arbitrary energy momentum tensor can be embedded, at least
locally, in a solution of the 5D Einstein field equations in
vacuum. Because of this, the stress-energy may be a 4D
manifestation of the embedding geometry. Physically, the
background metric there employed describes a 5D extension of an
usual de Sitter spacetime. Inflationary cosmology can be recovered
from a 5D vacuum\cite{nos,nosb,nosc}. Other version of 5D gravity,
which is mathematically similar, is the membrane theory. In this
theory gravity propagates freely on the 5D bulk and the
interactions of particles are confined to a 4D hypersurface called
"brane"\cite{rs,rsb,rsc}. Both versions of 5D general relativity
are in agreement with observations.

In this work we study both, particle and graviton production
after a phase transition occurred after inflation. To make it, we
shall consider the solutions obtained at the end of inflation from
a 5D vacuum on an effective 4D hypersurface described by a de
Sitter metric. After it, we shall use the Bogoliubov formalism to
calculate the modes solution for the fields after a phase
transition to a radiation dominated epoch. The topic here studied
is very important, because the case of inflaton decay into further
inflatons may be of interest for dark matter searches. Following
their production, the inflatons would decouple from the rest of
matter\cite{kls}. If the inflatons were given a tiny mass, then
these bosons could serve as a natural candidate for the missing
dark matter\cite{kaiser}.

\section{Induced Matter and Embeddings}

We consider a 5D manifold (${\cal M}, g_{ab}$) with a coordinate
system $x\equiv \left\{x^a\right\}$. We are interested in a 5D
theory of gravity on which we define a 5D vacuum, such that the
first variation of the action is $^{(5)} \delta{\cal I}= ^{(5)}
\delta{\cal I}_E+^{(5)} \delta{\cal I}_M$, where
\begin{equation}\label{ac}
^{(5)} \delta{\cal I} = \frac{1}{2} \int d^5x\, \sqrt{|g|}\,\delta
g^{ab}\,\left\{\frac{\hat{G}_{ab}}{8\pi G}+ \hat{T}_{ab} \right\}.
\end{equation}
The first term is the variation of the gravitational Einstein
action $^{(5)} {\cal I}_E$ and the second one is the variation of
the matter action $^{(5)} {\cal I}_M$. Here, $G$ is the
gravitational constant, $g$ is the determinant of the covariant
tensor metric $g_{ab}$\footnote{In this work $a,b$ run from $0$ to
$4$ and Greek letters rum from $0$ to $3$.} and $\hat{ R}$ is the
5D Ricci scalar on the metric. The energy-momentum tensor of
matter, $ \hat{T}_{ab}$ is defined from the variation of the
matter action $^{(5)} {\cal I}_M$ under a change of the metric,
and will be considered as null to describe the 5D apparent vacuum.
The Einstein equations on the 5D manifold ${\cal M}$ (we use
$c=\hbar=1$ units)
\begin{equation}
\hat{G}^a_{\,\,b} = -8 \pi G\, \hat{T}^a_{\,\,b},
\end{equation}
where $\hat{G}^a_b$ and $\hat{T}^a_b$ are respectively the
Einstein and Energy Momentum tensors on ${\cal M}$. Furthermore,
we can define a scalar function $l(x)$, which represents the
foliation of the higher-dimensional manifold. We shall consider
that the extra coordinate is space-like. Each hypersurface,
$\Sigma_l$, is considered as a 4D Lorentzian spacetime. We denote
$n_a$ as the normal vector to the hypersurface
$\Sigma_l$\footnote{In what follows we shall denote the covariant
derivative on the 5D hypersurface as $\nabla_a \,(..)$ and the
covariant derivative on $\Sigma_l$ with a semicolon: $(..)_{;a}$.}
\begin{equation}
n_a = -\nabla_a l_,
\end{equation}
such that the expression $n_a n^a =-1$ normalizes $n^a$. Now we
can define a coordinate system $y\equiv \{y^a\}$ on $\Sigma_l$.
The basis vectors are
\begin{equation}
e^a_{\alpha} = \frac{\partial x^a}{\partial y^{\alpha}}, \qquad
n_a e^a_{\alpha} =0.
\end{equation}
These objects can be used to project 5D tensors (for instance, the
metric tensor) into 4D ones (which lives on the $\Sigma_l$
hypersurface)
\begin{equation}
e^a_{\alpha} e^b_{\beta} \, g_{ab} = e^a_{\alpha} e^b_{\beta}
\,(h_{ab} +n_a n_b)=h_{\alpha\beta}.
\end{equation}

\subsection{Einstein tensor on $\Sigma_l$}

The extrinsic curvature $K_{\alpha\beta}$ of the 4D hypersurface
$\Sigma_l$ is a symmetric 2-range tensor given by the derivative
of the induced metric in the normal direction to $\Sigma_l$
\begin{equation}
K_{\alpha\beta} = e^a_{\alpha} e^{b}_{\beta} \nabla_a n_b.
\end{equation}
If now we consider an alternative coordinate system on ${\cal M}:
\{y^{\alpha},l\}$, such that
\begin{displaymath}
dx^a = e^a_{\alpha} \, dy^{\alpha} + l^a dl,
\end{displaymath}
where the vector tangent to lines with constant $y^{\alpha}$ can
be decomposed into the sum of a part tangent to $\Sigma_l$, and a
part normal to $\Sigma_l$
\begin{displaymath}
l^a =  N^{\alpha} e^a_{\alpha} + n^a,
\end{displaymath}
such that $l^a \partial_a l =1$. The 4D vector $N^{\alpha}$ is the
shift vector, which describes how the $y^{\alpha}$ coordinate
system changes as we move from a given $\Sigma_l$ hypersurface to
another. Finally, the 5D line element $dS^2$ can be written as
\begin{equation}
dS^2 = h_{ab} dx^a dx^b = h_{\alpha\beta} \left(dy^{\alpha} +
N^{\alpha} dl\right) \left(dy^{\beta} + N^{\beta} dl\right) -
dl^2,
\end{equation}
which, for a constant foliation $dl=0$, reduces to
\begin{equation}
ds^2 = h_{\alpha\beta}\, dy^{\alpha} dy^{\beta}.
\end{equation}
In this work we are interested in dealing with a 5D Riemann-flat
metric, so that the Ricci tensor is null
\begin{equation}\label{ua}
\hat{R}_{ab} =0.
\end{equation}
On each $\Sigma_l$ hypersurface, the Gauss-Codazzi equations
are\footnote{For a 5D Riemann-flat metric, are
$R_{\alpha\beta\gamma\delta} = 2 K_{\alpha [\delta}
K_{\gamma]\beta}$, and $K_{\alpha [\beta; \gamma]}=0$.}
\begin{eqnarray}
&& \hat{R}_{abcd} \,e^a_{\alpha} e^b_{\beta} e^c_{\gamma}
e^d_{\delta} = R_{\alpha\beta\gamma\delta} - 2 K_{\alpha [\delta}
K_{\gamma]\beta}, \\
&& \hat{R}_{mabc} n^m e^a_{\alpha} e^b_{\beta} e^c_{\gamma} = 2
K_{\alpha [\beta; \gamma]}.\label{ub}
\end{eqnarray}

If we use the expression for the 5D Ricci tensor
\begin{equation}\label{uc}
\hat{R}_{ab} = \left(h^{\mu\nu} e^m_{\mu} e^n_{\nu} - n^m
n^n\right) \hat{R}_{ambn},
\end{equation}
we obtain the contractions of (\ref{ua})
\begin{eqnarray}
&&\hat{R}_{ab} e^a_{\alpha} e^b_{\beta} =0, \nonumber \\
&& \hat{R}_{ab} e^a_{\alpha} n^b =0, \nonumber \\
&& \hat{R}_{ab} n^a n^b =0. \label{uu}
\end{eqnarray}
Using (\ref{uc}) into (\ref{uu}), and making use of (\ref{ub}), we
obtain the expressions
\begin{eqnarray}
&& R_{\alpha\beta} =E_{\alpha\beta} +K_{\alpha}^{\mu}
\left(K_{\beta\mu} - K\, h_{\beta\mu}\right) , \label{r}\\
&& \left(P^{\alpha\beta}\right)_{;\alpha} =\left( K^{\alpha\beta} - h^{\alpha\beta} K \right)_{;\alpha}=0, \label{eea}\\
&& E_{\mu\nu} h^{\mu \nu}=0,
\end{eqnarray}
where $K = h^{\alpha\beta} K_{\alpha\beta}$, $E_{\alpha\beta}=
E_{\beta\alpha}$ and $E_{\alpha\beta} = \hat{R}_{manb} n^m
e^a_{\alpha} n^n e^{b}_{\beta}$. Notice that the equation
(\ref{eea}) means that the second rank (symmetric) tensor
$P^{\alpha\beta}$ is conserved on the 4D hypersurface $\Sigma_l$.
It is important to notice that the Ricci tensor on $\Sigma_l$,
related to the 5D Riemann flat metric $\hat{R}_{mnlo}=0$, is
given by
\begin{equation}
R_{\alpha\beta} =K_{\alpha}^{\mu} \left(K_{\beta\mu} - K\,
h_{\beta\mu}\right).
\end{equation}

Finally, the Einstein tensor $G^{\alpha\beta} = R^{\alpha\beta} -
g^{\alpha\beta}\, R/2$ on a given $\Sigma_l$, is
\begin{equation}\label{ein1}
G^{\alpha\beta} = E^{\alpha\beta} + K^{\alpha}_{\mu} P^{\mu\beta}
- \frac{1}{2} h^{\alpha\beta} K^{\mu\nu} P_{\mu\nu}.
\end{equation}
Notice that for a 4D metric induced from a 5D Riemann-flat metric,
the Einstein tensor only depends on the extrinsic curvature
\begin{equation}\label{ein}
G^{\alpha\beta} =K^{\alpha}_{\mu} P^{\mu\beta} - \frac{1}{2}
h^{\alpha\beta} K^{\mu\nu} P_{\mu\nu}.
\end{equation}

\subsection{Einstein equations on  $\Sigma_l$}

In this work, we are interested in describing the primordial
universe on a 4D hypersurface $\Sigma_l$. In order to make it, we
shall consider a massless single scalar field $\varphi(x^a)$,
which, we shall consider to describe the physical vacuum on the 5D
manifold ${\cal M}$. The kinetic Lagrangian corresponding to this
field is
\begin{equation}
{\cal L}_{\varphi}=  \frac{1}{2} \sqrt{\left|\frac{
 g}{g_0}\right|}\,   g^{ab} \varphi_{,a}
\varphi_{,b},
\end{equation}
where $g= l^8 \exp{6(N-4\psi)}$ is the determinant of the
covariant metric tensor and $g_0$ is a dimensional constant.
Hence, to describe the energy momentum tensor related to
$\varphi$, on ${\cal M}$, we set
\begin{equation}
-8 \pi G\, \hat{T}^{ab} = \frac{\partial {\cal
L}_{\varphi}}{\partial g_{ab}} - \frac{\partial}{\partial x^{m}}
\left[\frac{\partial {\cal L}_{\varphi}}{\partial \left(
g_{ab,m}\right)}\right].
\end{equation}
It is easy to demonstrate that the $\hat{T}^{ab}$ projected on the
hypersurface $\Sigma_l$ is
\begin{equation}\label{em}
T^{\alpha\beta} = \hat{T}^{ab} e^{\alpha}_a e^{\beta}_b.
\end{equation}
Finally, using (\ref{em}) and (\ref{ein}), we obtain the Einstein
equations on $\Sigma_l$ obtained from a Riemann-flat 5D vacuum
\begin{equation}\label{eqq}
\left.G^{\alpha\beta}\right|_l = \left.K^{\alpha}_{\mu}
P^{\mu\beta} - \frac{1}{2} h^{\alpha\beta} K^{\mu\nu}
P_{\mu\nu}\right|_l = e^{\alpha}_a e^{\beta}_b \left.
\,\left\{\frac{\partial {\cal L}_{\varphi}}{\partial g_{ab}} -
\frac{\partial}{\partial x^{m}} \left[\frac{\partial {\cal
L}_{\varphi}}{\partial \left(
g_{ab,m}\right)}\right]\right\}\right|_l,
\end{equation}
that describes the equations of motion for the fields on
hypersurfaces with constant $l$ (i.e., for constant foliations).
Notice that the equations of motion obtained from a Ricci-flat 5D
metric must include the Einstein tensor (\ref{ein1}), rather than
(\ref{ein}), in the left side of the Einstein equations
(\ref{eqq}).

\section{Revisiting the formalism for scalar metric fluctuations on a 5D Riemann-flat metric}

We consider the Riemann-flat background metric\cite{lebe}
\begin{equation}\label{s1}
dS^2_b=l^2 dN^2-l^2 e^{2N} dr^2 -dl^2,
\end{equation}
where $dr^2=\delta_{ij}dx^{i}dx^{j}$. Here, $x^i$ are the 3D
cartesian space-like dimensionless coordinates, $N$ is a
dimensionless time-like coordinate and $l$ is the space-like
non-compact extra coordinate, which has length units. The
non-perturbative metric fluctuations of the background metric
(\ref{s1}), are introduced in our analysis by the line element
introduced in\cite{abm}
\begin{equation}\label{s2}
dS^2 =  l^2 e^{2\psi} dN^2 - l^2 e^{2(N- \psi)}dr^2-dl^2,
\end{equation}
where the metric function $\psi(N,\vec{r},l)$ describes the
gauge-invariant metric fluctuations with respect to the background
metric. In order to describe a 5D physical vacuum, we shall
consider a massless and free test scalar field $\varphi=\varphi
(x^{\alpha},l)$ defined on (\ref{s2}). The dynamics of $\varphi$
can be derived from the action
\begin{equation}\label{s3}
{\cal I}={\int} d^4 x \    dl \sqrt{\left|\frac{g}{g_0}\right|}
\left( \frac{\hat{R}}{16\pi G}+ \frac{1}{2} g^{ab} \varphi_{,a}
\varphi_{,b} \right).
\end{equation}
Now we consider a semiclassical approximation for the 5D scalar
field $\varphi$ in the form
$\varphi(N,\vec{r},l)=\varphi_{b}(N,l)+\delta\varphi
(N,\vec{r},l)$, with $\varphi _{b}$ denoting the background part
of $\varphi$ and $\delta\varphi$ denoting the quantum fluctuations
of $\varphi$, which will be considered as very small. This is
consistent with a linear approximation for the scalar metric
fluctuations. Thus, a first-order approximation in the gauge
scalar fluctuations of the form $e^{\pm 2\psi}\simeq 1\pm 2\psi$,
will be sufficient in order to have a good description of these
fluctuations during inflation, in the present formalism. After
making this approximation, one find that the dynamics on 5D of
$\varphi_{b}(N,l)$ and  $\delta\varphi (N,\vec{r},l)$, are
\begin{eqnarray}
\frac{\partial^{2}\varphi_{b}}{\partial N^2} &+&
3\frac{\partial\varphi_{b}}{\partial N}
-\left[l^{2}\frac{\partial^{2}\varphi_b}{\partial l^2}+4l\frac{\partial\varphi _{b}}{\partial l}\right]=0,\label{x3}\\
\frac{\partial^{2}\delta\varphi}{\partial N^2}&+&
3\frac{\partial\delta\varphi}{\partial
N}-e^{-2N}\nabla_{r}^{2}\delta\varphi-\left[l^{2}\frac{\partial^{2}\delta\varphi}{\partial
l^2}+4l\frac{\partial\delta\varphi}{\partial l}\right] \nonumber
\\
&-& 2\psi\left[l^{2}\frac{\partial^{2}\varphi_{b}}{\partial l^2} +
4l\frac{\partial\varphi_{b}}{\partial l}\right]=0. \label{x22}
\end{eqnarray}
The expression (\ref{x3}) gives the dynamics on the background of
$\varphi(N,\vec{r},l)$, whereas Eq. (\ref{x22}) describes the
dynamics for the quantum fluctuations $\delta\varphi(N,\vec{r},l)$
in terms of the scalar metric fluctuations $\psi$ and the
background field $\varphi _{b}$. On the other hand, the expression
for the energy momentum tensor components on the background is
\footnote{We denote with $\bar{g}_{ab}$ the background tensor
metric and $\varphi_b$ the solution for $\varphi(x^a)$ on
$\bar{g}_{ab}$.}
\begin{equation}
-8 \pi G\, \hat{T}_{ab} = \left.\frac{\partial {\cal
L}_{\varphi}}{\partial g^{ab}} - \frac{\partial}{\partial x^{m}}
\left[\frac{\partial {\cal L}_{\varphi}}{\partial \left(
g^{ab}_{\,\,\,\,,m}\right)}\right]\right|_{g_{ab}=\bar{g}_{ab},\varphi
= \varphi_b},
\end{equation}
where the Lagrangian density corresponding to the inflaton field
is
\begin{equation}
{\cal L}_{\varphi}=  \frac{1}{2} \sqrt{\left|\frac{
 g}{g_0}\right|}\,   g^{ab} \varphi_{,a}
\varphi_{,b}.
\end{equation}

One can find the {\em linearized equation} which describes the
dynamics of the salar metric fluctuations $\psi$ on the linearized
fluctuating metric $dS^2= (1+2\psi) l^2 dN^2 - l^2 e^{2N}
(1-2\psi) dr^2 - d\psi^2$\cite{abm}:
\begin{eqnarray}
\frac{\partial^{2}\psi}{\partial N^2} &+ &
7\frac{\partial\psi}{\partial N} +
6\psi-e^{-2N}\nabla_{r}^{2}\psi-\frac{2}{3}\left[14l\frac{\partial\psi}{\partial
l}+4l^{2}\frac{\partial^{2}\psi}{\partial l^2}\right] \nonumber
\\
&=& -\frac{8\pi G}{3}\left[\frac{\partial\varphi_b}{\partial
N}\frac{\partial\delta\varphi}{\partial N} + \psi
l^{2}\left(\frac{\partial\varphi _b}{\partial l}\right)^{2} +
l^{2}\frac{\partial\varphi_{b}}{\partial
l}\frac{\partial\delta\varphi}{\partial l}\right], \label{x15}
\end{eqnarray}
where $\psi$ complies with the linearized non-diagonal Einstein
equations
\begin{eqnarray}\label{x10}
&& \frac{\partial^{2}\psi}{\partial x^{i}\partial
N}+\frac{\partial\psi}{\partial x^i}=
4\pi G\frac{\partial\varphi_b}{\partial N}\frac{\partial\delta\varphi}{\partial x^i},\\
\label{x11} && \frac{\partial^{2}\psi}{\partial l\partial
N}+2\frac{\partial\psi}{\partial l}=\frac{8\pi G}{3}
\left(\frac{\partial\delta\varphi}{\partial
N}\frac{\partial\varphi_b}{\partial
l}+\frac{\partial\varphi_b}{\partial N}
\frac{\partial\delta\varphi}{\partial l}\right),\\
\label{x14} && \frac{\partial ^{2}\psi}{\partial x^{i}\partial
l}=8\pi G\left(\frac{\partial\varphi_b}{\partial
x^i}\frac{\partial\delta\varphi}{\partial
l}+\frac{\partial\delta\varphi}{\partial
x^i}\frac{\partial\varphi_b}{\partial l}\right).
\end{eqnarray}
These equations provide us with the dynamics for both, the scalar field
and the 5D gauge-invariant metric fluctuations\cite{abm}.

\section{Dynamics of fields at the end of inflation on an effective 4D de Sitter background}

We consider the background metric (\ref{s1}). If we take a static
foliation $l=l_0=1/H_0$, with the transformations
\begin{eqnarray}
&& N = t/l_0 = H_0 t, \\
&& R = l_0 r = 1/H_0 r,
\end{eqnarray}
we obtain the effective 4D de Sitter background metric
\begin{equation}\label{w1}
dS^2 = dt^2 - e^{2 H_0 t} dR^2.
\end{equation}
Here, $H_0$ is a constant that represents the Hubble parameter.
For the relativistic point of view, this means that now we shall
move with penta-velocities\footnote{Latin indices denote 3D
spatial coordinates.}
\begin{equation}
u^t =1, \qquad u^i = u^l =0.
\end{equation}

The effective 4D action on $\Sigma _H$ derived from the 5D action
(\ref{s3}), reads
\begin{equation}\label{w8}
^{(4)}{\cal I}_{eff}=\int d^{4}x\,\sqrt{\left|\frac{^{(4)}
 g}{^{(4)} g_0}\right|} \left( \frac{^{(4)} R}{16\pi G}+
\frac{1}{2}g^{\mu\nu} \Phi_{,\mu} \Phi_{,\nu} + V \right),
\end{equation}
where $\Phi(t,\vec{r})\equiv\varphi(t,\vec{r},l)|_{l=H_0^{-1}}$ is
the massive scalar field induced on $\Sigma _H$. Furthermore,
$^{(4)} g$ is the determinant of the 4D induced metric, which in
the case of background metric (\ref{w1}) yields
$^{(4)}\bar{g}=\exp(6H_0t)$. Moreover, in the case of the
perturbed metric (\ref{s2}) gives $^{(4)}g=\exp[2(3H_0t-2\Psi)]$,
and $^{(4)}g_0$ is a dimensionless constant. The induced 4D
effective potential $V$ in the action (\ref{w8}), has the form
\begin{equation}\label{w9}
V = -\left.\frac{1}{2} g^{ll}
\left(\frac{\partial\varphi}{\partial l}\right)^2\right|_{N=H_0
t,R=r/H_0,l=l_0=1/H_0}\,.
\end{equation}
Given the quantum nature of the fields $\Phi$ and $\Psi$ it seems
convenient to use the quantization procedure. To do it, we impose
the commutation relations
\begin{equation}\label{w10}
\left[\Phi(t,\vec R), \Pi^0_{(\Phi)}(t,\vec R')\right] = i\,
\delta^{(3)}\left(\vec R-\vec R'\right),\quad \left[\Psi(t,\vec
R), \Pi^0_{(\Psi)}(t,\vec R')\right] = i\, \delta^{(3)}\left(\vec
R-\vec R'\right),
\end{equation}
where $\Pi ^{0}_{(\Psi)}$ and $\Pi^{0}_{(\Phi)}$ are the canonical
momentums for $\Phi$ and $\Psi$, respectively. We can derive the
commutation relations on the effective 4D background metric. If
the inflaton and metric fluctuations are small, these commutators
can be approximated to
\begin{eqnarray}
&& \left[\Phi(t,\vec R), \dot\Phi(t,\vec R')\right] \simeq i
\sqrt{\left|\frac{^{(4)} g_0}{^{(4)} \bar{g}}\right|}
\delta^{(3)}\left(\vec R-\vec R'\right), \label{w11} \\
&& \left[\Psi(t,\vec R),  \dot\Psi(t,\vec R')\right] \simeq i
\frac{4}{9} \pi G \sqrt{\left|\frac{^{(4)} g_0}{^{(4)}
\bar{g}}\right|} \delta^{(3)}\left(\vec R-\vec R'\right).
\label{w22}
\end{eqnarray}
In this section we shall examine the evolution of the background
inflaton field, jointly with the metric and inflaton fluctuations.

\subsection{Dynamics of the background field at the end of inflation}

The dynamics of the background inflaton dynamics on the effective
4D background (de Sitter) metric is described by the equation
(\ref{x3}) with $l=l_0=1/H$ and $N=H_0 \,t$. In what follows we
shall define $\Phi_{b}(t)=\varphi_{b}(N=H_0 t,l=l_0=1/H_0)$. After
making separation of variables in (\ref{x3}), we obtain
\begin{equation}\label{af1}
\frac{\partial^{2}\Phi _b}{\partial
t^2}+3H_0\frac{\partial\Phi_b}{\partial t}+H_0^{2}m^{2}\Phi_{b}=0,
\end{equation}
$m$ being a dimensionless separation constant. The background
Friedmann equation $G^t_{\,\,t} = -8\pi G \left<
T^t_{\,\,t}\right>$, on the effective 4D hypersurface is
\begin{equation}\label{af2}
\left.\left(\frac{\partial\varphi_b}{\partial
t}\right)^{2}+H_0^{2}\left(l\frac{\partial\varphi_b}{\partial
l}\right)^{2}\right|_{l=H_0^{-1}}= \left.\frac{3}{4\pi G}
\frac{1}{l^2}\right|_{l=H_0^{-1}}=\frac{3H_0^{2}}{4\pi G}.
\end{equation}
The complete solution for eqs. (\ref{af1}) and (\ref{af2}) is with
$m=0$, such that the inflaton field is a constant of time: $\Phi_b
= \sqrt{\frac{1}{12 \pi G}}$\cite{abm}.

However, at the end of inflation the second term in the equation
(\ref{af1}) can be neglected and that equation can be approximated
to
\begin{equation}\label{af3}
\frac{\partial^{2}\Phi _b}{\partial
t^2}+H_0^{2}m^{2}\Phi_{b}\approx 0.
\end{equation}
In what follows we shall consider the dynamics of the inflaton
field governed by the approximated equation (\ref{af3}) with the
effective 4D Friedman equation (\ref{af2}). We shall consider that
at the end of inflation the mass of the inflaton field is of the
order of the Hubble parameter. In particular, if we consider
$m=2$, the dynamics of the metric fluctuations described by
(\ref{x15}) is simplified and we obtain the following solution for
(\ref{af3})
\begin{equation}\label{af4}
\Phi _b (t)= \Phi_0 \cos(2 H_0 t+\delta),
\end{equation}
where $\Phi_0 = \frac{1}{4} \sqrt{\frac{3}{\pi G}}$ and $\delta$
is some constant of integration to be determined.

\subsection{Dynamics of metric fluctuations at the end of
inflation}

Using the result (\ref{af4}), and combining equations (\ref{x15})
to (\ref{x10}) over the 4D hypersurfaces $\Sigma
_H:l=l_0=H_0^{-1}$, we obtain an uncoupled equation for
$\Psi(t,\vec{r})$

\begin{equation}\label{af5}
\frac{\partial^2\Psi}{\partial t^2} + 3 H_0
\frac{\partial\Psi}{\partial t} + 2 H_0^2 \Psi \left( 1 - 3
\cos^2(2 H_0 t +\delta) + S^2 \tan(2 H_0 t +\delta)\right) - 3
H_0^2 e^{-2 H_0 t}\nabla_r ^2 \Psi =0,
\end{equation}
where $S^2= 4 k^2_l$, $k_l$ being a separation constant.

Following a canonical quantization process, the field $\Psi$ can
be expanded in Fourier modes
\begin{equation}\label{m4}
\Psi(t,\vec{R}) = \frac{1}{(2\pi)^{3/2}}\int d^{3}k
\left[a_{k}\,e^{i\vec{k}
\cdot\vec{R}}\xi_{k}(t)+a_{k}^{\dagger}\,e^{-i\vec{k}\cdot\vec{R}}\xi_{k}^{*}(t)\right],
\end{equation}
where the annihilation and creation operators $a_{k}$ and
$a_{k}^{\dagger}$, satisfy the usual commutation algebra
\begin{equation}\label{m5}
[a_{k},a_{k'}^{\dagger}]=\delta ^{(3)}(\vec{k}-\vec{k'}),\quad
[a_{k},a_{k'}]=[a_{k}^{\dagger},a_{k'}^{\dagger}]=0.
\end{equation}

Using the commutation relation (\ref{m5}) and the Fourier
expansion (\ref{m4}), we obtain the normalization condition for
the modes $\xi_{k}(t)$
\begin{equation}\label{m6}
\xi_{k}(t) \dot{\xi}^*_{k}(t) - \xi^*_{k}(t) \dot{\xi}_{k}(t) = i
\frac{4\pi G }{9}\left(\frac{a_0}{a}\right)^{3} H_0^3,
\end{equation}
where $a(t)= e^{H_0t}$ is the scale factor in the 4D metric and
$a_{0}= e^{H_0t_0}$, being $t_{0}$ the cosmic time at the end of
inflation. Inserting (\ref{m4}) in (\ref{af5}), we obtain that the
modes $\xi_{k}(t)$ satisfy
\begin{equation}\label{m7}
\frac{\partial^2\xi_{k}}{\partial t^2} + 3 H_0
\frac{\partial\xi_{k}}{\partial t} + 2 H_0^2 \xi_{k} \left( 1 - 3
\cos^2(2 H_0 t +\delta) + \frac{3}{2} k^2 e^{-2 H_0 t} + S^2
\tan(2 H_0 t +\delta)\right)=0.
\end{equation}
Lets us consider the phase $\delta$ so that at the end of
inflation $2 H_0 t +\delta \approx \pi$. In this case, the
equation (\ref{m7}) leads to

\begin{equation}\label{m8}
\frac{\partial^2\xi_{k}}{\partial t^2} + 3 H_0
\frac{\partial\xi_{k}}{\partial t} + 2 H_0^2 \xi_{k} \left(
\frac{3}{2} k^2 e^{-2 H_0 t} -2 \right)=0.
\end{equation}

By means of the equation (\ref{m6}) and choosing the Bunch-Davies
vacuum condition, the normalized solution of (\ref{m8}),
is
\begin{equation}\label{m9}
\xi_{k}(t) = \frac{\pi
H_0}{3}\sqrt{G}\left(\frac{a_0}{a}\right)^{3/2} {\cal
H}^{(2)}_{\nu}[x(t)],
\end{equation}
where ${\cal H}^{(2)}_{\nu}[x(t)]$ is the second kind Hankel
function, $\nu =5/2$ and $x(t)= \sqrt{3} \: k \: e^{-H_0t}$.

\subsection{Dynamics of the inflaton field fluctuations at the end
of inflation}

Now, employing again the equations (\ref{x15}) to (\ref{x10}) over
the 4D hypersurfaces $\Sigma _H:l=l_0=H_0^{-1}$ and incorporating
the solution (\ref{af4}), we obtain that the 4D quantum
fluctuations for the inflaton field are determined by the
equation\cite{abm}

\begin{equation}\label{m10}
\frac{\partial^{2}\delta\Phi}{\partial
t^2}+3H_0\frac{\partial\delta\Phi}{\partial t} - H_0^2
e^{-2H_0t}\nabla_{r}^{2}\delta\Phi - H_0^2 \left. \left[ l^2
\frac{\partial^2 \delta \varphi}{ \partial l^2} + 4 l
\frac{\partial \delta \varphi}{ \partial l} \right]
\right|_{l=H_0^{-1}} + 8 H_0^2 \Psi \Phi_b=0,
\end{equation}

Performing a Fourier expansion of
$\delta\Phi$ in the form
\begin{equation}\label{m11}
\delta\Phi(t,\vec R) = \frac{1}{(2\pi)^{3/2}} {\int} d^3K \left[
a_{K} e^{i\vec{K}\cdot\vec R} \eta_{K}(t) + a^{\dagger}_{K}
e^{-i\vec{K}\cdot\vec R} \eta^*_{K}(t)\right],
\end{equation}
where the annihilation and creation operators $a_{K}$ and
$a_{K}^{\dagger}$ satisfy the commutation algebra given by
(\ref{m5}). Inserting (\ref{m10}) and (\ref{m4}) in (\ref{m9}), it leads to
\begin{equation}\label{m12}
\frac{\partial^2\eta_{K}}{\partial t^2} + 3 H_0
\frac{\partial\eta_{K}}{\partial N} + H_0^2 K^2 e^{-2H_0t} - H_0^2
\left. \left[ l^2 \frac{\partial^2 \eta_{K}}{ \partial l^2} + 4 l
\frac{\partial \eta_{K}}{ \partial l} \right] \right|_{l=H_0^{-1}}
+ 8 H_0^2 \xi_{K} \Phi_b=0.
\end{equation}

The previous equation is inhomogeneous, non separable and presents
a source term. The general solution will be the solution of the
homogeneous equation plus a particular solution

\begin{equation}\label{af6}
\eta_{K}(t)= \left[ f(t) \cos{(k \sqrt{3} e^{-H_0 t})} + g(t)
\sin{(k \sqrt{3} e^{-H_0 t})} \right] + \frac{e^{3 H_0
(t_0-t)}}{2} \sqrt{\frac{\pi}{H_0}} \: {\cal
H}^{(2)}_{\mu}\left[x(t)\right],
\end{equation}
where $f(t)$ and $g(t)$ are polinomial functions of $e^{\frac{H_0
t}{2}}$, and ${\cal H}^{(2)}_{\mu}\left[x(t)\right]$ is the second
kind Hankel function with $\mu=\frac{\sqrt{9-4 M^2}}{2}$ and
$x(t)=k e^{-H_0t}$. Here, we have used both, the Bunch -
Davies vacuum and the normalization conditions
\begin{equation}\label{aaf5}
\dot{\eta_h}_{K}^{*}{\eta_h}_{K}-\dot{\eta_h}_{K}{\eta_h}_{K}^{*}=i\left(\frac{a_0}{a}\right)^{3},
\end{equation}
for the Fourier modes of the homogeneous solution.

Towards the end of inflation one can neglect the lower terms of
$f(t)$ and $g(t)$, to take only those with higher order in
$e^{\frac{H_0 t}{2}}$. The term with the Hankel function can be
neglected too, so that we obtain
\begin{equation}\label{af7}
f(t)= i g(t) = \frac{40 H_0}{3\mu (25-4\mu^2)}
\sqrt{\frac{\sqrt{3}}{2 K^5}} \: e^{\frac{3}{2} H_0 t_0}\:
e^{\frac{5}{2}H_0 t}.
\end{equation}
Finally, the approximated solution for the modes of the inflaton
field fluctuations at the end of inflation can be rewritten as
\begin{equation}\label{af8}
\eta_{K}(t)= f(t) e^{-i k \sqrt{3}\:e^{-H_0t}},
\end{equation}
where $f(t)$ is given by (\ref{af7}).

\section{Particle Creation after a phase transition, after inflation}

To consider the particle creation lets first rewrite the solution
(\ref{af8}) in terms of the conformal time in a spatially flat 4D
Friedmann-Robertson-Walker cosmology $d\tau \equiv a^{-1} dt =
e^{-H_0t} dt$. Furthermore, we obtain that the modes of the
inflaton fluctuations are
\begin{equation}\label{af9}
\eta_{K}(\tau)= f(\tau) e^{i k \sqrt{3}\: H_0 a_0 \tau},
\end{equation}
and their derivatives with respect to $\tau$, which we denote with
a prime, are
\begin{equation}\label{af9a}
\eta'_{K}(\tau)= f(\tau) e^{i k \sqrt{3}\: H_0 a_0 \tau} \left( i K \sqrt{3} H_0 a_0 - \frac{5}{2 \tau} \right),
\end{equation}
where
\begin{equation}\label{af10}
f(\tau)= \frac{  40 \: H_0 \: e^{\frac{3}{2} H_0 t_0}}{3\mu
(25-4\mu^2)} \sqrt{\frac{\sqrt{3}}{2 K^5}} \: (-H_0
a_0\tau)^{-\frac{5}{2}}.
\end{equation}
In order to calculate the number of particles produced for each
mode ($N_k$), we match the solutions obtained at the end of
inflation, with those of the long-wavelength modes during a phase
transition from inflation to the radiation-dominated era. To
calculate this last solutions, we make use of the Bogolyubov
transformation (\cite{6}). For the long-wavelength modes we can
approximate the phase transition to be instantaneous, occurring at
some time $\tau_0$. So, we have that
\begin{equation}\label{transition1}
\begin{array}{ccl}
  $for$ \:\:\: & \:\:\:\tau < \tau_0: \:\:\: & \:\:\: a(\tau)= - \frac{1}{H \tau}, \\
  $for$ \:\:\: & \:\:\: \tau > \tau_0: \:\:\: & \:\:\: a(\tau)= - a_0 (\tau -
  \widetilde{\tau}),
\end{array}
\end{equation}
where $\widetilde{\tau}=\tau_0 - (H^2 \tau_0)^{-1}$. For
convenience, we set $a(\tau_0)=a_0$, so we obtain that
$\tilde{\tau}=0$, $H_0=a_0^{-1}$ and $\tau_0=-1$. After the
transition, the Hubble parameter is no longer constant, $H\equiv
H(\tau)$, and we get:

\begin{equation}
\begin{array}{ccl}
  $for$ \:\:\: & \:\:\:\tau < \tau_0: \:\:\: & \:\:\: H(\tau)= H_0, \\
  $for$ \:\:\: & \:\:\: \tau > \tau_0: \:\:\: & \:\:\: H(\tau)=
  \frac{H_0}{\tau^2}.
\end{array}
\end{equation}
In what follows we shall compute the number of particles and
gravitons in the radiation dominated era, which takes place after
a phase transition from the inflationary epoch.

\subsection{Particle creation from inflaton field fluctuations in the radiation dominated era}

In order to calculated the number of created particles after a
phase transition, we calculate the modes of the inflaton field
fluctuations taking into account the Bogolyubov transformations

\begin{equation}
\eta_K(\tau>\tau_0) = \frac{H_0 \tau}{\sqrt{2 \sqrt{3}\:K}} \left[\alpha_K e^{-i K \sqrt{3}\: \tau^{-1}} + \beta_K e^{i K \sqrt{3}\: \tau^{-1}} \right],
\end{equation}
so that their temporal derivatives are
\begin{equation}
\eta'_K(\tau>\tau_0) = \frac{H_0}{\sqrt{2 \sqrt{3}\:K}} \left[\alpha_K e^{-i K \sqrt{3}\: \tau^{-1}} \left( 1
+ \frac{i\sqrt{3} K}{\tau} \right) + \beta_K e^{i K \sqrt{3}\: \tau^{-1}} \left( 1 - \frac{i\sqrt{3} K}{\tau} \right) \right].
\end{equation}
To calculate the coefficients $\alpha_K$ and $\beta_K$, we use the
continuity of $\eta_K$ and $\eta'_K$ at $\tau = \tau_0 =-1$, so
that
\begin{equation}
\beta_K = a_0 \,f(\tau_0) \,\sqrt{2 K \sqrt{3}}\left( 1 -\frac{7
i}{2 K\sqrt{3}} \right),
\end{equation}

\begin{equation}
\alpha_K = 2\,a_0 \,f(\tau_0) \,\sqrt{2 K \sqrt{3}}\left(\frac{7
i}{2 K\sqrt{3}}-1  \right) e^{- 2 i K \sqrt{3}},
\end{equation}

\noindent where

\begin{equation}
f(\tau_0)= \frac{  40 \: H_0 \: e^{\frac{3}{2} H_0 t_0}}{3\mu
(25-4\mu^2)} \sqrt{\frac{\sqrt{3}}{2 K^5}}.
\end{equation}

Finally, by using $N_K =\left|\beta_K \right|^2$ we obtain the
number of created particles with a wavenumber $K$
\begin{equation}
N_K = 2\, a_0^2 \left| f(\tau_0) \,\right|^2 K \sqrt{3} \left( 1 +
\frac{29}{12  K^2} \right).
\end{equation}

To estimate numerically the number of particles created after the
phase transition on cosmological scales (small wavenumbers), we
shall consider that the number of e-folds, $N_e$, at the end of
inflation is at least\footnote{This means that we are considering
that the phase transition occurs abruptly after inflation ends.}:
$N_e>H_0 t_0 \sim 60$. The value of $K$ depends upon the details
of the transition between inflation and the radiation dominated
epoch. For long-wavelength modes (on cosmological super Hubble
scales), the transition can be considered as abrupt. In general
one expects that $K<1$\cite{vi}. For instance, if we take $0.01 <K
< 1$ and $H_0 \sim 6 \times 10^{-6}\: M_p$\footnote{The
gravitational constant $G = M_p^{-2}$.}, with $\mu (25-4\mu^2)
\simeq 24$\footnote{This is the value corresponding to a scale
invariant power spectrum (with $\mu=3/2$), for the inflaton field
fluctuations on cosmological scales.}, we obtain
\begin{equation}\label{pc}
 5.5 \times  10^{78} < N_K < 5.64 \times 10^{90},
\end{equation}
which, in cases where the wavenumber is small, is a number
sufficiently important to agree with the desired values for
entropy ${\cal S}
>10^{89}$.

\begin{figure*}
\includegraphics[height=15cm]{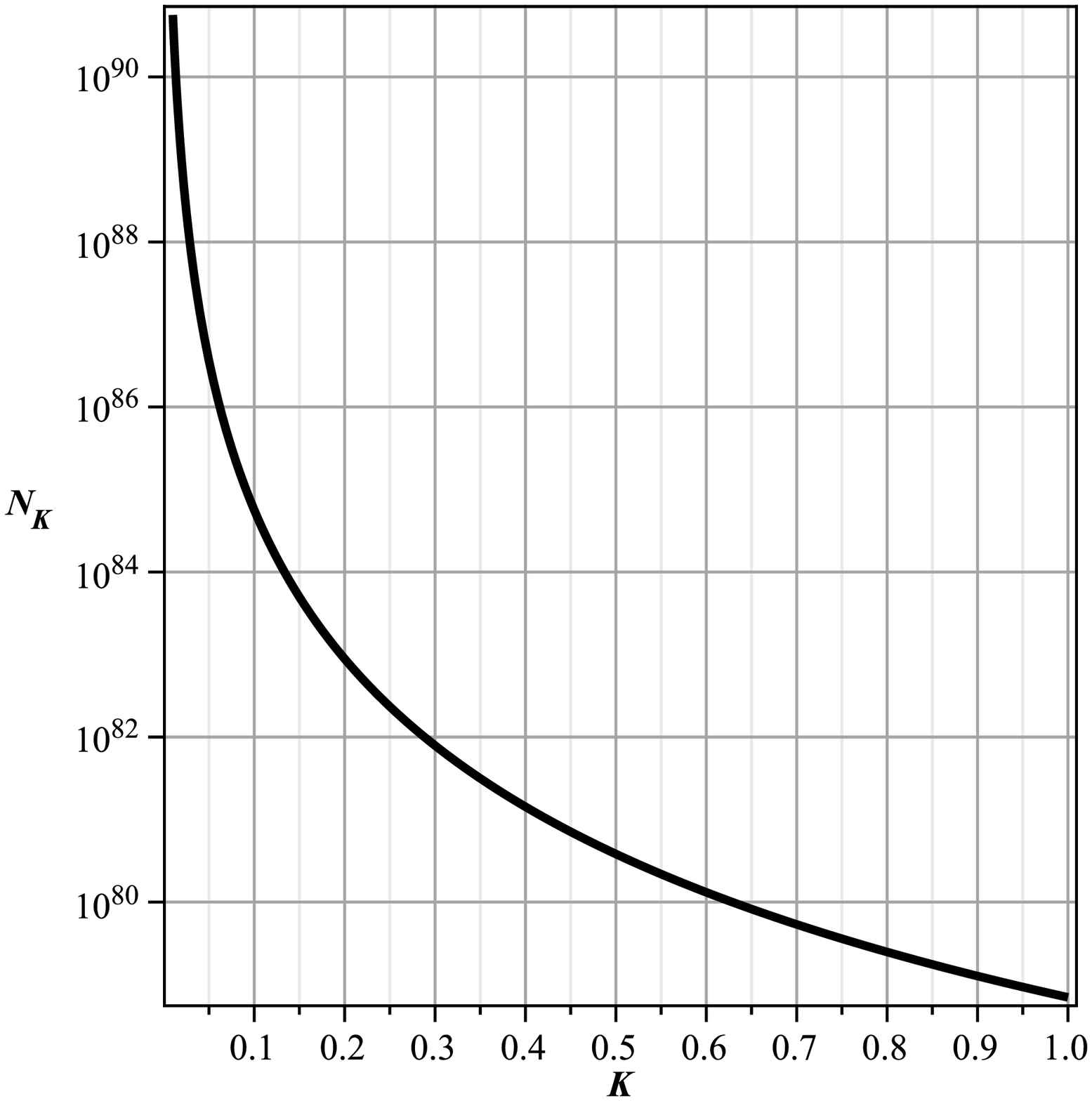}\caption{\label{figura1} Number of created particles after
inflation $N_K$, with respect to the wavenumber $0.01<K<1$.}
\end{figure*}

In the figure (\ref{figura1}) was plotted the number of particles
created for $0.01 <K < 1 $. Notice that the number of created
particles $N_K$ increases dramatically for cosmological scales.

\subsection{Graviton creation from scalar metric fluctuations}

We can also estimate the number of gravitons per mode generated by
the metric fluctuations at the end of inflation. Again we first
must rewrite the solution of the metric fluctuations modes in
conformal time $\tau$. The modes for the metric fluctuations
(\ref{m9}) can be expressed
\begin{equation}
\xi_{K}(\tau) = \mathcal{C}(\tau) e^{-i K \sqrt{3} H_0 a_0 \tau},
\end{equation}
so that the conformal time derivative is
\begin{equation}
\xi'_{K}(\tau) = \left( \mathcal{C}'(\tau) - i K \sqrt{3} H_0 a_0
\right) e^{-i K \sqrt{3} H_0 a_0 \tau}.
\end{equation}
where the function $\mathcal{C}(\tau)$ and its conformal time
derivative are given by
\begin{equation}
\mathcal{C}(\tau) = -\frac{H_0}{3} \sqrt{\frac{2 \pi G}{\sqrt{3}
K^5}} e^{\frac{3}{2} H_0 t_0}\left( -H_0 a_0 K^2 \tau + i K
\sqrt{3} +\frac{1}{H_0 a_0 \tau} \right),
\end{equation}
\begin{equation}
\mathcal{C}'(\tau) = \frac{1}{3} \sqrt{\frac{2 \pi G}{\sqrt{3} K}}
e^{\frac{3}{2} H_0 t_0} H_0^2 a_0 \left( 1 +\frac{1}{H_0^2 a_0^2
K^2 \tau^2} \right).
\end{equation}
Let us consider again the phase transition (\ref{transition1}),
with $a(\tau_0)=a_0$, so that we have $\tilde{\tau}=0$,
$H_0=a_0^{-1}$ and $\tau_0=-1$. After the phase transition, the
metric fluctuations modes can be written as

\begin{equation}
\xi_{K}(\tau>\tau_0) = \frac{\tau}{\sqrt{2 \sqrt{3}\:K}} \left[A_K
e^{-i K \sqrt{3}\: \tau^{-1}} + B_K e^{i K \sqrt{3}\: \tau^{-1}}
\right],
\end{equation}
so that
\begin{equation}
\xi'_{K}(\tau>\tau_0) = \frac{1}{\sqrt{2 \sqrt{3}\:K}} \left[A_K e^{-i K \sqrt{3}\: \tau^{-1}}
\left( 1 + \frac{i\sqrt{3} K}{\tau} \right)
+ B_K e^{i K \sqrt{3}\: \tau^{-1}} \left( 1 - \frac{i\sqrt{3} K}{\tau} \right) \right].
\end{equation}

To obtain the coefficients $A_K$ and $B_K$ we use the continuity
of $\xi_{K}$ and $\xi'_{K}$ at moment of the phase transition,
$\tau=\tau_0=-1$, and we obtain
\begin{eqnarray}
A_K &= &i \frac{2 H_0}{3 K^3} \sqrt{\frac{\pi G}{3}}
e^{\frac{3}{2} H_0 t_0}  +\sqrt{\frac{K \sqrt{3}}{ 2}}, \\
B_K &= & e^{2i K\sqrt{3}} \left\{ e^{\frac{3}{2} H_0 t_0}
\frac{\sqrt{\pi G} H_0}{3 K^2} \left[ \left( K^2 -2 \right) + i K
\sqrt{3} \left( 1 - \frac{2}{3 K^2}\right) \right]\right.
\nonumber \\
&-&\left. \sqrt{\frac{K \sqrt{3}}{2}} \right\}.
\end{eqnarray}

\begin{figure*}
\includegraphics[height=15cm]{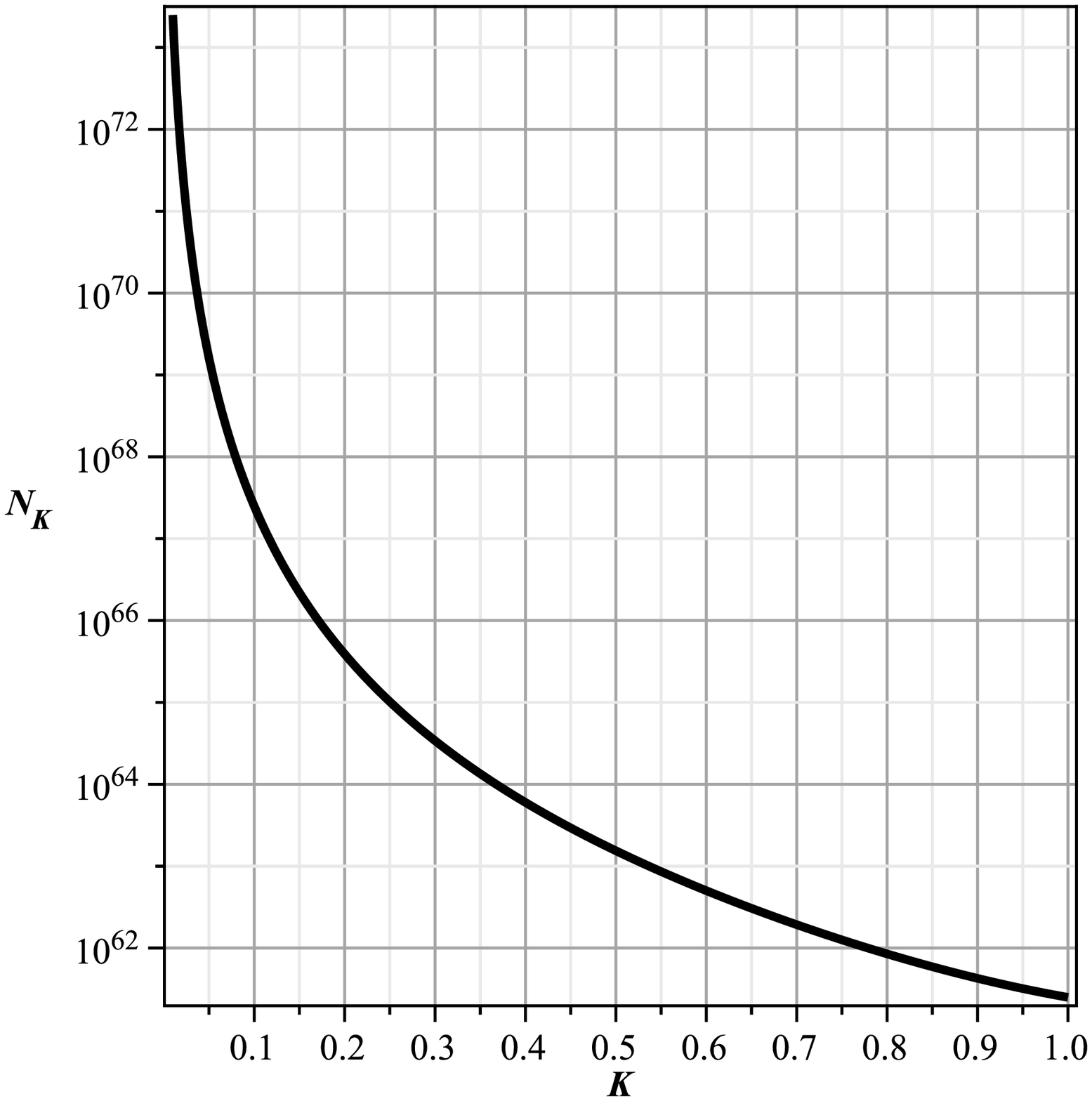}\caption{\label{figura2} Number of gravitons created after
inflation $N^{(g)}_K$, with respect to the wavenumber $0.01 < K <
1$.}
\end{figure*}

Using again $N^{(g)}_K =\left|B_K \right|^2$, we can obtain the
number of gravitons generated after the phase transition
\begin{eqnarray}
N^{(g)}_K &=& e^{3 H_0 t_0}\frac{\pi G  H_0^2}{9} \left( 1 -
\frac{1}{K^2} + \frac{4}{3 K^6} \right) - e^{\frac{3}{2} H_0 t_0}
\frac{\sqrt{2\pi G \sqrt{3}} H_0}{3} \left( K^{\frac{1}{2}} - 2
K^{-\frac{3}{2}}\right) \nonumber \\
&+& \frac{K \sqrt{3}}{2}
\end{eqnarray}
Notice that the third term inside the first bracket is the
dominant one. Now, if we consider the values of the parameters at
the end of inflation, we can calculate the approximate number of
gravitons created after the phase transition. In order to estimate
a numerical value, we set $H_0 t_0 > 60$ and $K \simeq 0.01$, so
that we obtain
\begin{equation}
N^{(g)}_K > 2.5 \times 10^{73},
\end{equation}
which is a number very small with respect to the number of
particles created (\ref{pc}) after the phase transition. In the
figure (\ref{figura2}) was plotted the number of gravitons created
$N^{(g)}_K$, which also increases dramatically on larger
cosmological scales.

\section{Final Comments}

We have studied particle and graviton production after an abrupt
phase transition from inflation to the radiation dominated era. We
have considered the Bogoliubov formalism to calculate the modes
solution for the fields after a phase transition to a radiation
dominated epoch. Our approach is different from others developed in
\cite{kaiser}, because here resonance effects are not important
for the inflaton and metric fluctuations. Furthermore, the
solutions for the modes of the inflaton fluctuations and the
scalar fluctuations of the metric were obtained using a recently
developed formalism which is related to the Induced Matter theory
of gravity. During inflation, such approach makes possible the
simultaneous description of the coupled dynamics of the inflaton
field fluctuations with the gauge-invariant fluctuations of the
metric. After inflation this solutions can be matched with those
of the radiation dominated epoch to obtain the Bogoliubov
coefficients. The results here obtained are very interesting,
because the number of created particles after inflation results to
be greater than $10^{90}$ at cosmological scales, and increases
dramatically as $K^{-6}$. Furthermore, $N_K$ is very sensitive to
the number of e-folds during inflation. This is due to the fact that
$N_K$ increases as $a(t)^3$. In what respect to the number of
gravitons created after inflation, the results are qualitatively
similar to those of created particles. However, the number of
gravitons created are, at least, of the order of $10^{73}$, which
is much smaller ($10^{-17}$ times) than $N_K$. All these results
were obtained using the values $H_0\,t_0 =60$, with $H_0 =
10^{-6}\, M_p$, and wavenumbers $K < 10^{-2}$.

\section*{Acknowledgements}

\noindent The authors acknowledge UNMdP and CONICET Argentina for
financial support.

\bigskip


\begin{thebibliography}{99}
\bibitem{1} A. H. Guth, Phys. Rev. D 23 (1981) 347.
\bibitem{2} D. H. Lyth and A. Riotto, Phys. Rept. {\bf 314}: 1 (1999).
\bibitem{3} A.D. Linde, Physics and Inflationary
Cosmology, Harwood, Chur, Switzerland, 1990.
\bibitem{4} A. R. Liddle and D. H. Lyth, Cosmological inflation and large-scale
structure, Cambridge University Press, 2000.
\bibitem{5} M. Bellini, {\em et al}. Phys. Rev. {\bf D54}: 7172 (1996).
\bibitem{5b} E. W. Kolb, S. Matarrese, A. Notari and A. Riotto, Mod. Phys.
Lett. {\bf A20}:2705 (2005).
\bibitem{cal} Esteban Calzetta. Phys. Rev. {\bf D44}:3043(1991).
\bibitem{calb} A. A. Grib and Y. V. Pavlov. Grav. Cosmol. {\bf 11}: 119
(2005).
\bibitem{chao} A. D. Linde. Phys. Lett. {\bf B129}: 177(1983).
\bibitem{ci} Vadim Kuzmin and Igor Tkachev, {\bf JETP} Lett. {\bf
68}: 271 (1998).
\bibitem{wi} Gert Aarts and Anders Tranberg, Phys. Lett. {\bf
B650}: 65 (2007).
\bibitem{fi} Mauricio Bellini, Nuovo Cim. {\bf
B117}: 653 (2002).
\bibitem{bi} Hassan Firouzjahi and Salomeh Khoeini-Moghaddam, {\bf JCAP
1102}: 012 (2011).
\bibitem{rad} A. D. Dolgov and A. D. Linde. Phys. Lett. {\bf
B116}: 329 (1982).
\bibitem{radb} L. F. Abbott, E. Farhi and M. B.
Wise. Phys. Lett. {\bf B117}: 29 (1982).
\bibitem{radc} D. V.
Nanopoulos, K. A. Olive, M. Srednicki. Phys. Lett. {\bf B127}: 30
(1983).
\bibitem{radd} Y. Shtanov, J. H. Traschen, R. H.
Brandenberger. Phys. Rev. {\bf D51}: 5438 (1995).
\bibitem{rade}
L. Kofman, A. Linde, A. A. Starobinsky. Phys. Rev. Lett. {\bf 76}:
1011 (1996).
\bibitem{kaiser} D. I. Kaiser, Phys. Rev. {\bf D53}, 1776 (1996).
\bibitem{abm} M. Anabitarte, M. Bellini, J. E. Madriz Aguilar, Eur. Phys. J. {\bf C65}:295
(2010).
\bibitem{campbell} J. E. Campbell, {\em A course of Differential
Geometry} (Charendon, Oxford, 1926);\\
L. Magaard, {\em Zur einbettung riemannscher Raume in
Einstein-Raume und konformeuclidische Raume}. (PhD Thesis, Kiel,
1963).
\bibitem{campbellb} S. Rippl, C. Romero, R. Tavakol, Class.
Quant. Grav. {\bf 12}: 2411 (1995).
\bibitem{campbellc} F. Dahia, C. Romero, J. Math. Phys.{\bf 43}: 5804 (2002).
\bibitem{campbelld} F. Dahia, C. Romero, Class. Quant. Grav. {\bf 22}: 5005 (2005).
\bibitem{nos} M. Bellini. Phys. Lett. {\bf B609}: 187 (2005).
\bibitem{nosb} J. E. Madriz Aguilar and M. Bellini. Phys. Lett. {\bf B619}: 208
(2005). \bibitem{nosc} M. Anabitarte and M. Bellini. J. Math.
Phys.{\bf 47}: 042502 (2006).
\bibitem{rs} L. Randall and R. Sundrum, Mod. Phys. Lett. {\bf
A13}:2807 (1998). \bibitem{rsb} L. Randall and R. Sundrum, Phys.
Rev. Lett. {\bf 83}:4690 (1999).
\bibitem{rsc} L. F. P. da Silva,
J. E. Madriz Aguilar, Mod. Phys. Lett. {\bf A23}:1213 (2008).
\bibitem{kls} L. Kofman, A. Linde and A. A. Starobinsky, Phys.
Rev. Lett. {\bf 73}: 3195 (1994).
\bibitem{lebe} D. S. Ledesma and M. Bellini, Phys. Lett. {\bf B581}:1 (2004).
\bibitem{6} N. D. Birrell and P. C. W. Davies, {\em Quantum Fields
in Curved Space}. Cambridge University Press, Cambridge, England
(1982).
\bibitem{vi} T. Damour and A. Vilenkin. Phys. Rev. {\bf D53}: 2981
(1996).
\end{thebibliography}
\end{document}